\documentclass[12pt]{article}
\usepackage{epsfig}
\usepackage{amssymb}
\usepackage{amsmath}
\usepackage{graphicx}

\newcommand{\eqdef}{\stackrel{\text{def}}{=}}
\newcommand{\n}{\nonumber \\}

\newcommand{\ignore}[1]{}
\newcommand{\Romannumeral}[1]{\uppercase\expandafter{\romannumeral#1}}
\newcommand{\I}{\text{\Romannumeral{1}}}
\newcommand{\II}{\text{\Romannumeral{2}}}

\allowdisplaybreaks[4]

\def\be{\begin{equation}}
\def\ee{\end{equation}}
\def\ba{\begin{array}{c}}
\def\ea{\end{array}}

\newcommand{\bea}{\begin{eqnarray}}
\newcommand{\eea}{\end{eqnarray}}

\newcommand{\Title}[1]{{\baselineskip=26pt
  \begin{center} \Large \bf #1 \\ \ \\ \end{center}}}
\newcommand{\Author}{\begin{center}
  \large \bf  Ryu Sasaki${}^1$  and Miloslav Znojil${}^2$ \end{center}}
\newcommand{\Address}{\begin{center}
    ${}^1$ Faculty of Science, Shinshu University,
     Matsumoto 390-8621, Japan\\
      e-mail: ryu@yukawa.kyoto-u.ac.jp\\[12pt]
${}^2$ Nuclear Physics Institute ASCR, 
Hlavn\'{\i} 130, 250 68 \v{R}e\v{z}, Czech Republic\\
e-mail: znojil@ujf.cas.cz
   \end{center}}
\newcommand{\Accepted}[1]{\begin{center}
{\small \sf Published in:}\quad   {\large \sf #1}
  \end{center}}
\begin{document}

\thispagestyle{empty}

\Title{One-dimensional Schr\"{o}dinger equation with non-analytic potential
$V(x)= -g^2\exp (-|x|)$ and its exact Bessel-function solvability}

\Author

\Address

\vspace{0.5cm}

\begin{abstract}
Exact solvability (ES) of one-dimensional quantum
potentials $V(x)$ is a vague concept. We propose that beyond its most
conventional range the ES status should be attributed also to many
less common interaction models for which the wave functions remain
piecewise proportional to special functions. The claim is supported
by constructive analysis of a toy model $V(x)= -g^2\exp (-|x|)$. The
detailed description of the related bound-state and scattering
solutions of Schr\"{o}dinger equation is provided in terms of Bessel
functions which are properly matched in the origin.
\end{abstract}
\subsection*{Keywords:}

non-analytic potentials; bound states; reflection and transmission;
orthogonality theorems;
associated Hamiltonians;
supersymmetry;

\subsection*{PACS}

 03.65.Ge Solutions of wave equations: bound states\\
 02.30.Gp Special functions

\Accepted{Journal of Physics  {\bf A49} (2016) Nr.44 445303 (12pp)}

\newpage

\section{Introduction}

In the current literature on quantum mechanics an almost
disproportionate amount of attention is being attracted by the so called
exactly solvable (ES) models of dynamics. After a more detailed
inspection the concept gets split into several subcategories. The
main divide may be spotted between the ES property of the piece-wise
constant ``point'' interactions (cf., e.g., Ref.\,\cite{Albeverio})
and the ES property of a family of polynomially solvable analytic
potentials (cf., e.g., one of their lists in Ref.\,\cite{Cooper}
and the recent developments of the {\em exceptional} and the
{\em multi-indexed orthogonal polynomials} in Refs.\,\cite{gomez}--\cite{os25}).

The resulting point-or-analytic classification of the ES models is
not too satisfactory, for several reasons. In what follows we intend
to emphasize that one of these reasons is that the respective
construction techniques (i.e., the matching of wave functions in the
case of point interactions and the closed-form constructions of wave
functions in the case of analytic interactions) are independent and
could be combined. Thus, in place of the  point-or-analytic divide
one encounters a non-empty third domain of the possible coexistence of the
two approaches.

In the language of mathematics the new, intermediate ES domain will
inherit some weaknesses of the former, elementary-function ES domain
which admits non-polynomial wave functions and in which the related
determination of spectra is partially numerical. Only the latter,
analytic-solvability domain remains, strictly speaking,
non-numerical and characterized not only by the polynomiality of the
wave functions but also by the presence of various inherent symmetries
 in combination with a comparatively low degree of
flexibility -- {\it pars pro toto} let us mention the frequent
``shape invariance'' property  and/or the supersymmetry-representing
relations between the different analytic ES potentials
\cite{Cooper}. In contrast, the essence of the former,
point-interaction approach to the one-dimensional Schr\"{o}dinger
equation
 \begin{equation}
-\frac{d^2}{dx^2}\psi(x)+V(x)\psi(x) =E\,\psi(x),\ \ \ \  x\in
(-\infty,\infty)
 \label{SE}
 \end{equation}
may be seen in a suitable {\it ad hoc} split of the real line of
coordinates $x$ into subintervals,
\begin{equation*}
 (-\infty,\infty)= (-\infty,a_1)
\bigcup (a_1,a_{2})\bigcup \ldots \bigcup (a_K, \infty)\,.
\end{equation*}
Inside
these subintervals the potential function $V(x)$ is being chosen
constant (cf., e.g., the square-well model Nr.\,26 on p.\,52 in
\cite{Fluegge} with $K=4$ or several other models of this type in
\cite{Constantinescu}) or at least sufficiently elementary -- cf.,
e.g., the $K=2$ Coulomb + square-well model Nr.\,28 on p.\,64 in
\cite{Constantinescu} or the most recent symmetrized Morse-potential
short-range-interaction one-dimensional model of Ref.\,\cite{symorse}
with $K=1$ and $a_1=0$.

In practice, the piecewise-nonconstant potentials $V(x)$ are rarely
treated as ES (cf., e.g., \cite{papone} or the model Nr.\,28 in
 \cite{Constantinescu}
once more). Even in the above-mentioned methodical
studies~\cite{symorse,papone} it has been revealed that from the purely
numerical point of view the traditional ``shooting and bracketing''
algorithm seems more efficient than the direct use of the available
(viz., confluent hypergeometric function) piecewise analytic and
asymptotically correct representations of the individual bound-state
wave functions $\psi_n(x) \in L^2(\mathbb{R})$.

Recently, a fairly persuasive counterargument against the resulting
scepticism has been provided by Ishkhanyan \cite{Ishkh1}. In a way
which we now perceive as disproving the discouraging experience of
Refs.\,\cite{symorse,papone} he took into consideration a new $K=1$ model
with $V(x) \sim 1/\sqrt{|x|}$ (and, in our present language, with
$a_1=0$), he constructed its asymptotically correct wave functions
$\psi_n(x) \in L^2(\mathbb{R})$ (again, in a closed form derived
from the confluent hypergeometric functions) and, finally, he
succeeded in showing that the matching condition (in the origin)
appears numerically extremely friendly.

The Ishkhanyan's conclusions may be perceived as a gem of a new
paradigm, recommending an immediate extension of the standard ES
domain of the use of matching techniques which were, up to now,
virtually exclusively applied just to the piecewise constant
potentials. In parallel, the Ishkhanyan's experience may be also
perceived as a decisive encouragement of transition to the
infinite-series versions of special functions. Indeed, a transition
to non-truncated, non-polynomial confluent hypergeometric wave
functions is also one of the key characteristics of his
``screened-Coulomb'' model with $V(x) \sim 1/\sqrt{|x|}$.
See also the most recent Ref.\cite{afterIshkhanyan} in this context.

In certain sense these innovative observations may be perceived as
weakening the well known exclusivity of the confluent hypergeometric
functions in elementary quantum mechanics (well sampled by their
applications in supersymmetric quantum mechanics \cite{Cooper}).
Indeed, the essence of their popularity may be traced back to their
ability of termination (to Laguerre polynomials). Naturally, there
exist several other special functions {\em without} such a
degeneracy option. In the new paradigmatic framework, their possible
use in solving Eq.~(\ref{SE}) should certainly be thoroughly
reconsidered at present, therefore.

We shall recall here the Bessel functions as one of the most natural
illustrative examples. A serendipitous additional merit of these
functions is that they might provide the piecewise analytic general
solutions of Eq.~(\ref{SE}) in the case of various piecewise
exponential potentials such that $V(x) = \alpha_j\exp \beta_j x$ for $x \in
(a_j,a_{j+1})$. In what follows we shall simplify the problem
(choosing $K=1$ and $a_1=0$) and we shall bring a number of
arguments supporting the effective ES status of the model.

\section{Bound states}
\label{II}

\subsection{Problem setting}
\label{set}

Our one-dimensional Schr\"odinger equation (\ref{SE}) will be
considered  with potential
\begin{equation}
V(x) =-g^2\exp(-|x|),\quad x\in(-\infty,\infty), \quad g>0\,.
\label{pot}
\end{equation}
The model may admit, first of all, the discrete and non-degenerate and finitely many
bound-state energies $E_m=-\kappa^2_m$ at $m=0,1,\ldots,M$. 
See the fourth paragraph in Appendix A in this connection. 
The
corresponding eigenfunctions must be normalizable, $\psi_m(x) \in
L^2(\mathbb{R})$. Since the potential is parity invariant,
$V(-x)=V(x)$, the eigenfunctions are also parity invariant,
\begin{equation}
\psi_m(-x)=(-1)^m\psi_m(x)\,.
\end{equation}
According to the conventional oscillation theorems \cite{Hille} the
subscript $m$ counts the {\em nodes} of the eigenfunctions.
Moreover, we may only consider the (say, positive) half-axis of
$x>0$, with
\begin{equation}
\text{even parity}:\quad\psi'_{2n}(0)=0,\qquad \text{odd
parity}:\quad \psi_{2n+1}(0)=0\,, \label{bc}
\end{equation}
i.e., with the eigenfunctions constrained by the parity-dependent
boundary condition in the origin.

\subsection{Eigenfunctions}

Let us introduce an auxiliary function $\rho(x)$,
\begin{equation}
\rho(x)\eqdef 2g e^{-|x|/2},\quad
\frac{d\rho(x)}{dx}=\text{sign}(-x)\frac{\rho(x)}{2}, \label{rprop}
\end{equation}
which maps $[0,\infty)$ and $(-\infty,0]$ to $(0,2g]$. With this
function, the Schr\"odinger equation for the discrete spectrum
$E=-\kappa^2$ is rewritten as the equation for Bessel functions,
\begin{equation}
\psi(x)=\phi(\rho(x))\ \Rightarrow
\frac{d^2\phi(\rho)}{d\rho^2}
+\frac1{\rho}\frac{d\phi(\rho)}{d\rho}
+\left(1-\frac{4\kappa^2}{\rho^2}\right)\phi(\rho)=0.
\end{equation}
The general solutions are obtained as linear combinations of two
types of Bessel functions. In the MAPLE-inspired notation \cite{MAPLE} we may use
either the negative-sign ansatz,
\begin{equation}
\psi(x)=A\, J(-2\kappa,\rho(x))+B\, Y(-2\kappa,\rho(x)),\quad
A,B\in\mathbb{R},
\end{equation}
or its positive-sign alternative,
\begin{equation}
\psi(x)=A\, J(2\kappa,\rho(x))+B\, Y(2\kappa,\rho(x)),\quad
A,B\in\mathbb{R}. \label{psol}
\end{equation}
These solutions are to be constrained by the appropriate asymptotic
boundary conditions and by the matching of their logarithmic
derivatives in the origin.

\subsection{Eigenvalues}
\label{discr}
Let us discuss the positive order solutions \eqref{psol} in some
detail. First let us note that, for $\kappa\in\mathbb{R}_{>0}$,
function
$J(2\kappa,\rho)$ is {\em non-singular} at $\rho=0$ or $x=\infty$;
\begin{equation}
J(2\kappa,\rho)\simeq \left(\frac{\rho}2\right)^{2\kappa}\times
O(1),\quad \rho\simeq0,\qquad J(2\kappa,2ge^{-x/2})\simeq
g^{2\kappa}e^{-\kappa x}\times O(1),\quad x\to+\infty,
\label{asymp+}
\end{equation}
whereas functions  $J(-2\kappa,\rho)$ and $Y(\pm 2\kappa,\rho)$ are {\em
singular} at $\rho=0$ or $x=\infty$.
Thus, 
wave functions satisfying the matching conditions \eqref{bc} in the
origin can be easily found. Since $x=0$ corresponds to a regular
point $\rho=2g$ of the Bessel functions we have, for the {\em even}
wave functions,
\begin{align}
\psi^{(e)}(x)&=A^{(e)}(\kappa,g)\,
J(2\kappa,\rho(x))+B^{(e)}(\kappa,g)\, Y(2\kappa,\rho(x)),\\
A^{(e)}(\kappa,g)&\eqdef -Y'(2\kappa,2g)
=-\left(Y(2\kappa-1,2g)-\frac{\kappa}{g}Y(2\kappa,2g)\right),\\
B^{(e)}(\kappa,g)&\eqdef
J'(2\kappa,2g)=J(2\kappa-1,2g)-\frac{\kappa}{g}J(2\kappa,2g).
\label{ecomb}
\end{align}
In deriving the second expressions for $A^{(e)}$ and $B^{(e)}$,  the
following two relations for the general Bessel functions
$Z(\nu,\rho)$ were employed,
\begin{equation}
\frac{dZ(\nu,\rho)}{d\rho}=\frac12\left(Z(v-1,\rho)-Z(\nu+1,\rho)\right),\quad
Z(\nu-1,\rho)+Z(\nu+1,\rho)=\frac{2\nu}{\rho}Z(\nu,\rho)\,.
\end{equation}
For the {\em odd} wave functions the result is equally
straightforward,
\begin{align}
\psi^{(o)}(x)&=A^{(o)}(\kappa,g)\,
 J(2\kappa,\rho(x))+B^{(o)}(\kappa,g)\, Y(2\kappa,\rho(x)),\\
A^{(o)}(\kappa,g)&\eqdef -Y(2\kappa,2g),\\
B^{(o)}(\kappa,g)&\eqdef J(2\kappa,2g). \label{ocomb}
\end{align}
Thus, one obtains the {\em eigenfunctions} whenever the coefficient
of the singular term vanishes,
\begin{align}
\psi_{2n}(x)&=J(2\kappa_{2n},\rho(x))=\left\{
\begin{array}{cc}
J(2\kappa_{2n},2ge^{-x/2})  &   x\ge0   \\[2pt]
J(2\kappa_{2n},2ge^{x/2})   &   x\le0
\end{array}
\right., \  \ E_{2n}=-\kappa_{2n}^2, \quad n=0,1,\ldots,
\label{evenfun}\\
\psi_{2n+1}(x)&=\text{sign}(x)J(2\kappa_{2n+1},\rho(x))=\left\{
\begin{array}{cc}
J(2\kappa_{2n+1},2ge^{-x/2})&   x\ge0   \\[2pt]
-J(2\kappa_{2n+1},2ge^{x/2})&   x\le0
\end{array}
\right. ,\\
&\hspace{50mm} E_{2n+1}=-\kappa_{2n+1}^2, \quad n=0,1,\ldots
\label{oddfun}
\end{align}
and whenever our solutions are matched, properly, in the origin,
\begin{align}
\text{even:}&\quad
J'(2\kappa_{2n},2g)=0=J(2\kappa_{2n}-1,2g)-\frac{\kappa_{2n}}{g}J(2\kappa_{2n},2g),\quad
n=0,1,\ldots,
\label{even}\\
\text{odd:}&\quad J(2\kappa_{2n+1},2g)=0,\quad n=0,1,\ldots\,.
\label{odd}
\end{align}
The latter relations may be  perceived as implicit definitions
of the eigenvalues,
 \be
  -g^2<E_0<E_1<E_2<\cdots<E_M <0\quad \Longleftrightarrow \quad
g>\kappa_0>\kappa_1>\kappa_2>\cdots > \kappa_M >0\,. \label{e-kappa}
 \ee
Finally, from the estimates \eqref{asymp+} of the behavior of our
Bessel functions $J(2\kappa,\rho)$ near $\rho=0$ we may deduce
that our eigenfunctions have  the correct physical asymptotics,
\begin{align}
\psi_{2n}(x)&\simeq \text{const.}\left\{
\begin{array}{cc}
e^{-\sqrt{-E_{2n}}x}&\quad \ x\to+\infty\\
e^{\sqrt{-E_{2n}}x}& \quad \  x\to-\infty
\end{array}
\right.,\\
\psi_{2n+1}(x)&\simeq \text{const.}\left\{
\begin{array}{cc}
e^{-\sqrt{-E_{2n+1}}x}& x\to+\infty\\
-e^{\sqrt{-E_{2n+1}}x}& x\to-\infty
\end{array}
\right..
\end{align}
In this manner we have established that our ``non-analytic
exponential well potential'' (\ref{pot}) is {\em exactly solvable}.
Moreover, the compact form of the solutions enables us  to
verify easily the mutual orthogonality between the even and odd
eigenfunctions. The evaluation of the normalization
coefficients from relations
\begin{align}
\int_0^{\infty}J(2\kappa_{2n},2ge^{-x/2})
J(2\kappa_{2m},2ge^{-x/2})dx\propto
\delta_{n\,m},
\label{orteven}\\
\int_0^{\infty}J(2\kappa_{2n+1},2ge^{-x/2})
J(2\kappa_{2m+1},2ge^{-x/2})dx\propto
\delta_{n\,m}, \label{ortodd}
\end{align}
is left to the readers.

\section{Scattering problem}
\label{scatt}

\begin{figure}[h]                     
\begin{center}                        
\epsfig{file=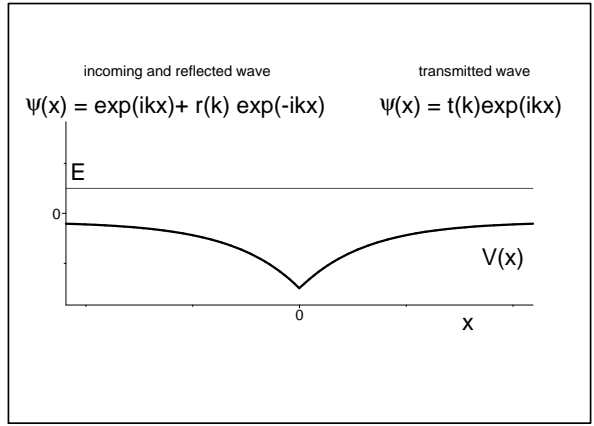,angle=270,width=0.5\textwidth}
\end{center}                        
\vspace{-2mm}\caption{An experimental arrangement of the scattering in one
dimension ($E=k^2$).
 \label{picee3wwww}}
\end{figure}

In the light of the results of the preceding section one may ask
whether our model is also exactly solvable at the positive energies
$E=k^2$, i.e., in the continuous-spectrum dynamical regime of
scattering. In the one-dimensional version of this regime the
experimentally testable information is carried by the reflection and
transmission amplitudes (cf. Fig.~\ref{picee3wwww}).

In order to determine the reflection amplitude $r(k)$ and the
transmission amplitude $t(k)$, let us solve  Schr\"odinger equation
$\mathcal{H}\psi_k(x)=k^2\psi_k(x)$ with positive
$k\in\mathbb{R}_{>0}$ and with the following boundary conditions:
\begin{align}
\psi_k(x)&\approx\left\{
\begin{array}{cl}
 e^{ikx} &   \quad x\to+\infty   \\
A(k) e^{ikx}+B(k)e^{-ikx}&  \quad  x\to-\infty
\end{array}
\right.,
\label{full}\\[4pt]
r(k)&=\frac{B(k)}{A(k)},\qquad t(k)=\frac1{A(k)}.
\end{align}
For $\rho\to0$, the asymptotics of the relevant  Bessel functions are
\begin{equation}
\rho\to0,\quad J(2ik,\rho)\simeq
\left(\frac{\rho}2\right)^{2ik},\quad J(-2ik,\rho)\simeq
\left(\frac{\rho}2\right)^{-2ik}.
\end{equation}
This leads to the following asymptotics in variable $x$,
\begin{align}
x\to+\infty,& \quad J(2ik,\rho(x))\simeq g^{2ik}e^{-ikx},\quad
J(-2ik,\rho(x))\simeq g^{-2ik}e^{ikx},\\
x\to-\infty,& \quad J(2ik,\rho(x))\simeq g^{2ik}e^{ikx},\quad \ \
J(-2ik,\rho(x))\simeq g^{-2ik}e^{-ikx}.
\end{align}
According to \eqref{full} we have to connect
\begin{align}
x\le0: \quad &A(k)g^{-2ik}J(2ik,\rho(x))+ B(k)g^{2ik}J(-2ik,\rho(x)),\\
x\ge0: \quad & g^{2ik}J(-2ik,\rho(x)),
\end{align}
at $x=0$. That is, these functions have to be equal at $x=0$
together with the first derivatives. We obtain
\begin{align}
&A(k)g^{-2ik}J(2ik,2g)+ B(k)g^{2ik}J(-2ik,2g)=g^{2ik}J(-2ik,2g),\\
&A(k)g^{-2ik}J'(2ik,2g)+ B(k)g^{2ik}J'(-2ik,2g)=-g^{2ik}J'(-2ik,2g).
\end{align}
This set of two linear equations can be readily solved,
\begin{align}
\begin{pmatrix}
A(k)\\
B(k)
\end{pmatrix}
&=\frac1{W}
\begin{pmatrix}
g^{2ik}J'(-2ik,2g)&-g^{2ik}J(-2ik,2g)\\
-g^{-2ik}J'(2ik,2g)&g^{-2ik}J(2ik,2g)
\end{pmatrix}\times
\begin{pmatrix}
g^{2ik}J(-2ik,2g)\\
-g^{2ik}J'(-2ik,2g)
\end{pmatrix}\\
&=\frac1{W}
\begin{pmatrix}
2g^{4ik}J'(-2ik,2g)J(-2ik,2g)\\
-\left(J'(2ik,2g)J(-2ik,2g)+J'(-2ik,2g)J(2ik,2g)\right)
\end{pmatrix}\,.
\label{AB}
\end{align}
In this formula we abbreviated
\begin{align}
W\eqdef J(2ik,2g)J'(-2ik,2g)-J'(2ik,2g)J(-2ik,2g)=i \sinh(2 \pi
k)/(\pi g)=i|W| \label{pure}
\end{align}
and used one of the Lommel's relations (\S3.12 Eq.(2) of \cite{watson})
\begin{align}
J(\nu,\rho)J'(-\nu,\rho)-J'(\nu,\rho)J(-\nu,\rho)=-2\sin(\nu\pi)/(\pi\rho)\,.
\end{align}
As long as the second argument $2g$ of our Bessel functions is
always the same we may omit it and shorten our formulae,
$J(\pm2ik,2g) \to J(\pm2ik)$. This yields the amplitudes in compact
form,
\begin{align}
r(k)&=B(k)/A(k)=-\frac{\left( J'(2ik)J(-2ik)+J(2ik)J'(-2ik)
\right)}{2J'(-2ik)J(-2ik)}\times g^{-4ik},\\
t(k)&=1/A(k)=\frac{W}{2J'(-2ik)J(-2ik)}\times g^{-4ik}.
\end{align}
For $k>0$ and $g>0$, it is straightforward to  verify that these
scattering amplitudes satisfy the {\em unitarity relations}
\begin{align}
r(k)t(k)^*+r(k)^*t(k)=0,\quad |r(k)|^2+|t(k)|^2=1,
\end{align}
in which the star $^*$ marks complex conjugation.

The proof proceeds as follows. Firstly, since $J(2ik)^*=J(-2ik)$ and
$J'(2ik)^*=J'(-2ik)$ we may conclude that $\left(
J'(2ik)J(-2ik)+J(2ik)J'(-2ik) \right)$ is real whereas $W$ in
\eqref{pure} is purely imaginary. The orthogonality
$r(k)t(k)^*+r(k)^*t(k)=0$ also holds. Thus, we obtain
\begin{align}
|r(k)|^2+|t(k)|^2&=\frac1{4J'(-2ik)J(-2ik)J'(2ik)J(2ik)}\n
&\qquad \times\left( \left( J'(2ik)J(-2ik)+J(2ik)J'(-2ik)
\right)^2+|W|^2\right)\,.
\end{align}
In other words we have to show that
\begin{align}
\left( J'(2ik)J(-2ik)+J(2ik)J'(-2ik)
\right)^2+|W|^2=4J'(2ik)J(2ik)J'(-2ik)J(-2ik)\,.
\end{align}
Fortunately, this is rather simple due to \eqref{pure},
\begin{align}
\text{l.h.s.}-\text{r.h.s.}&= \left( J'(2ik)J(-2ik)-J(2ik)J'(-2ik)
\right)^2+|W|^2=W^2+|W|^2=0.
\end{align}

In conclusion let us notice that in our model the zeros of $A(k)$ of
Eq.~\eqref{AB} (continued to the positive imaginary axis, $k\to
i\kappa$, $\kappa>0$) correspond, as they should, to the discrete
eigenvalues of the system. The demonstration is again easy: from
formula
\begin{align}
k\to i\kappa\qquad J'(-2ik,2g)J(-2ik,2g)\to
J'(2\kappa,2g)J(2\kappa,2g),
\end{align}
we see that  the zeros of the first factor $J'(2\kappa,2g)$
correspond to the even eigenstates \eqref{even} while those of the
second factor $J(2\kappa,2g)$ correspond to the odd eigenstates
\eqref{odd}.

\section{Associated Hamiltonians\label{v}}

\subsection{The Crum's sequence}

According to Crum \cite{crum}, to a one-dimensional Hamiltonian
$\mathcal{H}=\mathcal{H}^{[0]}$ with the eigensystems
$\{E_n,\psi_n(x)\}$, $n=0,1,\ldots$, a sequence of {\em
iso-spectral} Hamiltonian systems $\mathcal{H}^{[L]}$ $L=1,2,\ldots$,
is associated:
\begin{align}
\mathcal{H}^{[L]}\psi_n^{[L]}(x)&=E_n\psi_n^{[L]}(x),
\quad n=L,L+1,\ldots,\\
\mathcal{H}^{[L]}&\eqdef\mathcal{H}^{[0]}
-2\partial_x^2\log\left|\text{W}[\psi_0,\psi_1,\ldots,\psi_{L-1}](x)\right|,\\
\psi_n^{[L]}(x)&\eqdef\frac{\text{W}[\psi_0,\psi_1,\ldots,
\psi_{L-1},\psi_n](x)}{\text{W}[\psi_0,\psi_1,\ldots,\psi_{L-1}](x)},
\label{wronM}
\end{align}
in which the Wronskian of $n$-functions $\{f_1,\ldots,f_n\}$ is
defined by formula
\begin{align}
&\text{W}\,[f_1,\ldots,f_n](x)
  \eqdef\det\Bigl(\frac{d^{j-1}f_k(x)}{dx^{j-1}}\Bigr)_{1\leq j,k\leq n}.
  \label{wron}
  \end{align}
This result is obtained from a multiple application of the Darboux
transformations \cite{darboux}. One only has to adopt the lowest
eigenfunctions as the seed solutions $\{\psi_0,\psi_1,\ldots\}$. The
following properties of the Wronskians are instrumental:
\begin{align}
  &\text{W}[gf_1,gf_2,\ldots,gf_n](x)
  =g(x)^n\text{W}[f_1,f_2,\ldots,f_n](x),
  \label{Wformula1}\\
    &\text{W}[f_1(y),f_2(y),\ldots,f_n(y)](x)\n
&\qquad \qquad
  ={y'(x)}^{n(n-1)/2}\text{W}[f_1,f_2,\ldots,f_n](y),
  \label{Wformula0}\\
  &\text{W}\bigl[\text{W}[f_1,f_2,\ldots,f_n,g],
  \text{W}[f_1,f_2,\ldots,f_n,h]\,\bigr](x)\n
  &=\text{W}[f_1,f_2,\ldots,f_n](x)\,
  \text{W}[f_1,f_2,\ldots,f_n,g,h](x)
  \qquad(n\geq 0).
  \label{Wformula2}
\end{align}
By using the properties of the Wronskians
\eqref{Wformula1}--\eqref{Wformula0}, we can reduce the Wronskians
of $\{\psi_n(x)\}$ of Eq.~\eqref{wronM}  to the Wronskians of the
Bessel functions of the first kind, $\{J(2\kappa_n,\rho)\}$. For
example, we obtain:
\begin{align}
\text{W}[\psi_0,\psi_n](x)&=\left(\text{sign}(-x)\right)^{1+n}\frac{\rho}{2}
\cdot \text{W}[J(2\kappa_0,\rho),J(2\kappa_n,\rho)](\rho),\n
\text{W}[\psi_0,\psi_1,\psi_n](x)&=\left(\text{sign}(-x)\right)^{2+n}
\left(\frac{\rho}{2}\right)^3\!\cdot
\text{W}[J(2\kappa_0,\rho),J(2\kappa_1,\rho),J(2\kappa_n,\rho)](\rho),\n
\text{W}[\psi_0,\psi_1,\ldots,\psi_{L-1},\psi_n](x)&=
\left(\text{sign}(-x)\right)^{L+n}\left(\frac{\rho}{2}\right)^{L(L+1)/2}\cdot\n
&\quad\times
\text{W}[J(2\kappa_0,\rho),J(2\kappa_1,\rho),\ldots,J(2\kappa_{L-1},\rho),
J(2\kappa_n,\rho)](\rho).
\end{align}
The next-to-elementary structure of these relations indicates that 
the further systematic study of the $L$-th associated Hamiltonians
remains unexpectedly friendly and transparent. In particular,
we believe that 
the constructive solution of the related
scattering problem is so straightforward that it can be left
to the readers as an exercise.

\subsection{The breakdown of the shape invariance}

Let us consider the associated Hamiltonian systems
$\mathcal{H}^{[L]}$ with $L=1,2,\ldots$ corresponding to the
``non-analytic exponential well'' \eqref{pot}. Obviously they are
all {\em exactly solvable}. It is easy to see that the systems are
parity invariant:
\begin{align}
V^{[L]}(x)&\eqdef V(x)-2\partial_x^2\log
\left|\text{W}[\psi_0,\psi_1,\ldots,\psi_{L-1}](x)\right|,
\quad V^{[L]}(-x)=V^{[L]}(x),\\
\psi_n^{[L]}(-x)&=(-1)^{L+n}\psi_n^{[L]}(x).
\end{align}
By construction all the eigenfunctions $\{\psi_n(x)\}$ and their first derivatives $\{\psi'_n(x)\}$
are continuous. By using the Schr\"odinger equation, the second
derivatives in the Wronskians \eqref{wronM} are replaced by
\begin{align}
  \psi_n''(x)\longrightarrow
  \left(V(x)+\kappa^2_n\right)\psi_n(x),
 \end{align}
in which $V(x)\psi_n(x)$ is canceled by the multi-linearity of the
determinant. This applies to all the even order derivatives and the
Wronskian \eqref{wronM} contains the eigenfunctions and their first
derivatives only. Thus the eigenfunctions $\{\psi_n^{[L]}(x)\}$ of
the associated systems and their first derivatives are continuous.

Because of the parity, the orthogonality relations among the even
and odd eigenfunctions are trivial and those even-even and odd-odd
\begin{align}
\delta_{n\,m}\propto
(\psi_n^{[L]},\psi_m^{[L]})=\int_{-\infty}^{\infty}
\psi_n^{[L]}(x)\psi_m^{[L]}(x)dx
\end{align}
can be rewritten as those on the positive $x$-axis
\begin{align}
\delta_{n\,m}&\propto \int_{0}^{\infty}
\psi_{2n}^{[L]}(x)\psi_{2m}^{[L]}(x)dx,
\label{ortMeven}\\
\delta_{n\,m}&\propto \int_{0}^{\infty}
\psi_{2n+1}^{[L]}(x)\psi_{2m+1}^{[L]}(x)dx. \label{ortModd}
\end{align}

It is straightforward to evaluate
\begin{align}
-2\partial_x^2\log\psi_0(x)&=\frac{\rho(x)^2}{2}
-2\kappa_0^2+\frac{\rho(x)^2}{2}
\times\left(\frac{J'(2\kappa_0,\rho(x))}{J(2\kappa_0,\rho(x))}\right)^2,
\end{align}
which gives the potential of the first associated Hamiltonian
$\mathcal{H}^{[1]}$
\begin{align}
V^{[1]}(x)\eqdef V(x)-2\partial_x^2\log\psi_0(x)=
\frac{\rho(x)^2}{4}-2\kappa_0^2+\frac{\rho(x)^2}{2}
\times\left(\frac{J'(2\kappa_0,\rho(x))}{J(2\kappa_0,\rho(x))}\right)^2.
\label{V1}
\end{align}
If the potential of $V^{[1]}(x)$ \eqref{V1}  has the form (up to an
additive constant),
\begin{align*}
-f(g)^2e^{-|x|},
\end{align*}
with a certain function $f(g)$ of the parameter $g$, the system is
{\em shape invariant}, which is a sufficient condition for exact
solvability \cite{genden}. This simply cannot happen, since
$\frac{J'(2\kappa_0,\rho(x))}{J(2\kappa_0,\rho(x))}$ can never be a
constant. By construction, the potential of $\mathcal{H}^{[1]}$ is
non-analytic function of $x$ due to the $\rho(x)=2ge^{-|x|/2}$
dependence. This gives the {\em  negative answer} to the question
whether ``the solvability is due to the shape invariance''.

\section{Discussion}

One can  consult the textbook by Watson \cite{watson} to confirm that
the {\em zeros of Bessel functions $J(\nu,x)$ (i.e, 
in the notation of the book,  of $J_{\nu}(x)$) as functions of subscript $\nu$ for
fixed $x$ were not discussed in the book.} A few comments on this
problem were relocated here to a mathematical Appendix A. In it,
Theorems I and II could be interesting, {\it per se}, as they could
be very special cases of much more general theorems for complex $x$.

Another feature of our present results which is worth emphasizing is that
our one-dimensional model strongly resembles the spherical version
$V(r)=-g^2e^{-r/a}$ of the same potential in three or more
dimensions. In the s-wave sector the related radial Schr\"{o}dinger
equation is known to reduce to the Bessel equation. From our present
point of view
this recipe picks up, exclusively,
just the `odd' solutions of our present paper. Moreover, the
well known fact is that there is a minimal value of the
interaction strength $g$ which yields at least one discrete
eigenstate in the s-wave setting.
Indeed, one has the constraint
$g^2 a^2\ge j_{01}/4$, in which $j_{01}$ is the minimal positive
zero of $J_0(x)$, $j_{01}\simeq 2.405$.
After transition to
the full real line of coordinates and to the even solutions, the
surprising news
is that, in a
way proved in Appendix A, at least one discrete even-parity
eigenstate always exists for any nonvanishing coupling $g$.

We may summarize that our present model is exactly solvable
not only in the bound-state dynamical regime but also in the
continuous spectrum sector. We  succeeded in obtaining the
reflection and transmission amplitudes and in demonstrating,
constructively, that they satisfy the expected unitarity relations.
These constructions become more interesting when we analytically
continue  and reveal that in spite of the violation of the
analyticity of the potential (\ref{pot}) itself (in $x$), the poles
of the scattering amplitudes (in the complex planes of energies or
momenta) correspond to the discrete spectrum.

Last but not least, in a sharp contrast to the s-wave predecessors of
our present constructions the one-dimensional form of the present
problem also enabled us to recall the Crum's results and to point out
that the existence of the sequence of the associated Hamiltonian
systems should be perceived as a highly nontrivial result with,
perhaps, multiple future consequences. {\it Pars pro toto} we
emphasized here the non-shape invariance of this hierarchy of
models.
In this connection, let us remind the readers of another class of
 non-shape invariant and exactly solvable potentials, that is, the
 {\em reflectionless} potentials of Refs.~\cite{KM,refless}.

\subsection*{Acknowledgements}

The MZ's participation in the project was supported by the
Institutional Research Plan RVO61389005 and by GA\v{C}R Grant Nr.
16-22945S.


\section*{Appendix A. Orthogonality theorems}

The orthogonality theorems for the eigenfunctions of the original
``non-analytic exponential well'' \eqref{orteven}--\eqref{ortodd}
and for the associated systems \eqref{ortMeven}--\eqref{ortModd} can
be stated as simple theorems for Bessel functions.

In this Appendix we shall temporarily return to the textbook,
subscripted denotation $J_{\nu}(x)$ of the present Bessel functions
$J(\nu,x)$. Next, having fixed a positive number $x$, we shall
consider the Bessel functions of the first kind $J_{\nu}(x)$ and
$J'_{\nu}(x)\eqdef\frac{d}{dx}J_{\nu}(x)$ as functions of the order
$\nu$. This leads to the following {corollary}:  For fixed $x$,
there are finitely many ($M+1$) positive zeros of our interest ($M=\mathcal{N}+\mathcal{M}+1$),
\begin{align*}
J'_{\nu}(x):& \quad x>\lambda_0>\lambda_1>\cdots>\lambda_{\mathcal
N}>0,\quad J'_{\lambda_j}(x)=0,\\
J_{\nu}(x):& \quad x>\mu_0>\mu_1>\cdots>\mu_{\mathcal M}>0,\quad
J_{\mu_j}(x)=0\,.
\end{align*}
The oscillation theorem for the potential \eqref{pot} tells us that
these nodal zeros are {\em interlaced}, i.e.,
\begin{equation}
 x>\lambda_0>\mu_0>\lambda_1>\mu_1>\cdots>\mu_{\mathcal
 N-1}>\lambda_{\mathcal N}>0,
\end{equation}
or
\begin{equation}
  x>\lambda_0>\mu_0>\lambda_1>\mu_1>\cdots>\mu_{\mathcal
 N-1}>\lambda_{\mathcal N}>\mu_{\mathcal N}>0.
\end{equation}
This means that the number of the positive zeros  of $J_{\nu}(x)$ is equal or
smaller than its partner from $J'_{\nu}(x)$, i.e.,
$\mathcal{M}=\mathcal{N}$ or $\mathcal{M}= \mathcal{N}-1$.

In particular, one notices a slightly counterintuitive fact that a
nodeless even-parity ground state with $2\,{ \kappa}_{0}=\nu$ always
exists, irrespectively of the size of the strength $g=z/2$ of the
attraction. An independent proof of this observation is elementary:
the corresponding wave function
 $J_\nu(z)$
may be approximated by its first two terms,
\begin{equation*}
  {J}_\nu(z) \approx (z/2)^\nu/\Gamma(\nu+1)-(z/2)^{\nu+2}/\Gamma(\nu+2)+\ldots
\end{equation*}
so that the matching condition of Eq.~\eqref{even} yields
\begin{equation*}
 z^2=\frac{4\nu(\nu+1)}{\nu+2} \approx 2\nu + \ldots\,.
\end{equation*}

Next, by the change of variables $x\to\rho$, the orthogonality
relations between all of the eigenfunctions
\eqref{orteven}--\eqref{ortodd}
of the original system can be stated as follows. \\
{\bf Theorem \I}
\begin{align}
\text{even}:& \quad
\int_0^{x}J_{\lambda_j}(\rho)J_{\lambda_k}(\rho)\frac{d\rho}{\rho}=0,\qquad
j\neq k,
\label{theo1even}\\
\text{odd}:& \quad
\int_0^{x}J_{\mu_j}(\rho)J_{\mu_k}(\rho)\frac{d\rho}{\rho}=0,\qquad
j\neq k. \label{theo1odd}
\end{align}
Let us denote these two types of zeros by one consecutive sequence
($\{\nu_j\}$):
\begin{align*}
\nu_0\equiv \lambda_0,\ \nu_1\equiv \mu_0,\ \nu_2\equiv \lambda_1,\
\nu_3\equiv \mu_1,\ldots,.
\end{align*}
Likewise, the  orthogonality relations of the eigenfunctions
\eqref{ortMeven}--\eqref{ortModd} of the $L$-th associated
Hamiltonian system can be stated as\\
{\bf Theorem \II}
\begin{align}
&\text{even}:\int_0^{x}\frac{\text{W}[J_{\nu_0},\ldots,J_{\nu_{L-1}},
J_{\nu_{2n}}](\rho) \text{W}[J_{\nu_0},\ldots,J_{\nu_{L-1}},
J_{\nu_{2m}}](\rho)}
{\left(\text{W}[J_{\nu_0},\ldots,J_{\nu_{L-1}}](\rho)\right)^2}\rho^{2L-1}
d\rho=0,\quad n\neq m,
\label{theo2even}\\
&\text{odd}:\int_0^{x}\frac{\text{W}[J_{\nu_0},\ldots,J_{\nu_{L-1}},
J_{\nu_{2n+1}}](\rho) \text{W}[J_{\nu_0},\ldots,J_{\nu_{L-1}},
J_{\nu_{2m+1}}](\rho)}
{\left(\text{W}[J_{\nu_0},\ldots,J_{\nu_{L-1}}](\rho)\right)^2}\rho^{2L-1}
d\rho=0,\quad n\neq m. \label{theo2odd}
\end{align}
Theorem \I\ is the special case ($L=0$) of Theorem \II.

\section*{Note added}

After completing the paper, we became aware of a remark in \cite{erdelyi}
\begin{quote}
Investigations about the zeros $\nu_n$ of $J_\nu(z)$ regarded as a function of $\nu$, 
with fixed $z$ have been carried out by Coulomb (1936) \cite{coul}.  
They show that for positive real values of $z$, the $\nu_n$ are real and simple and asymptotically  near to negative integers.
\end{quote}
Coulomb's results were derived based on an indefinite integration formula 
of Bessel functions in Watson's textbook \cite{watson} (formula (13) on page 135, \S5.11)
\begin{align} 
 \int^z \mathcal{C}_{\mu}(kz)\overline{\mathcal{C}}_{\nu}(kz)\frac{dz}{z}&=
-\frac{kz\left\{\mathcal{C}_{\mu+1}(kz)\overline{\mathcal{C}}_{\nu}(kz)
-\mathcal{C}_{\mu}(kz)\overline{\mathcal{C}}_{\nu+1}(kz)\right\}}{\mu^2-\nu^2} \nonumber\\[-4pt]
&\quad \mbox{}+\frac{\mathcal{C}_{\mu}(kz)\overline{\mathcal{C}}_{\nu}(kz)}{\mu+\nu},
\end{align}
in which $\mathcal{C}_{\mu}$ and  $\overline{\mathcal{C}}_{\nu}$ are two arbitrary cylinder functions.
It is easy to derive Theorem I for the odd part (66) from the above formula 
by putting  $k=1$, $\mathcal{C}_{\mu}(z)\to J_\mu(z)$, $\overline{\mathcal{C}}_{\nu}(z)\to J_\nu(z)$
and integrating on $[0,z]$.
The even part of Theorem \I\ (65) can also be obtained in the same manner by noting,
in terms of (13), $J'_{\mu}(z)=0$\ 
$\Rightarrow J_{\mu-1}(z)=J_{\mu+1}(z)=\frac{\mu}{z}J_{\mu}(z)$.
As is clear from the above derivation,   Theorem I is valid for complex $z$, $\text{Re}(\mu)>0$, $\text{Re}(\nu)>0$.
However, for complex $z$, the oscillation theorem is not obvious and the validity of the 
interlacing relations (63), (64) is unclear. 

Let us emphasize, however, that in our case Theorem \I\ is not just definite integral formulas, 
but the {\em orthogonality relation} representing the exact solvability of potential (2),
which in turn leads to higher orthogonality relations, Theorem \II\ via Crum's sequence.
It is an interesting challenge  to see if Theorem \II\ is also valid for complex $z$. 

We found another paper \cite{conde} discussing the $\nu$ zeros of $J_{-\nu}(x)$.






\begin{thebibliography}{90}

\bibitem{Albeverio}
S. Albeverio, F. Gesztesy, R. Hoegh-Krohn and  H. Holden,
{\sl Solvable Models in Quantum Mechanics},
Springer, New York, (1988).


\bibitem{Cooper}
F. Cooper, A. Khare and U. Sukhatme,
``Supersymmetry and quantum mechanics,''
Phys. Rep. {\bf 251} (1995) 267-388.

\bibitem{gomez}
D.\,G\'{o}mez-Ullate, N.\,Kamran and R.\,Milson,
``An extension of Bochner's problem: exceptional invariant subspaces,''
J. Approx Theory {\bf 162} (2010) 987-1006,
{\tt arXiv:0805.\hspace{0pt}3376[math-ph]};
``An extended class of orthogonal polynomials defined by a
Sturm-Liouville problem,''
J. Math. Anal. Appl. {\bf 359} (2009) 352-367,
{\tt arXiv:0807.3939\hspace{0pt}[math-ph]}.

\bibitem{quesne}
C.\,Quesne,
``Exceptional orthogonal polynomials, exactly solvable potentials
and supersymmetry,''
J. Phys. {\bf A41} (2008) 392001,
{\tt arXiv:0807.4087[quant-ph]}.

\bibitem{os16}
S.\,Odake and R.\,Sasaki,
``Infinitely many shape invariant potentials and new orthogonal polynomials,''
Phys. Lett. {\bf B679} (2009) 414-417,
{\tt arXiv:0906.0142\hspace{0pt}[math-ph]}.

\bibitem{os25}
S.\,Odake and R.\,Sasaki,
``Exactly Solvable Quantum Mechanics and Infinite Families of
Multi-indexed Orthogonal Polynomials,"
Phys. Lett. {\bf B702} (2011) 164-170,
{\tt arXiv:\hspace{0pt}1105.0508[math-ph]}.


\bibitem{Fluegge}
S. Fl\"ugge, {\sl Practical Quantum Mechanics I},  Springer, Berlin, (1971).

\bibitem{Constantinescu}
F. Constantinescu and E. Magyari, {\sl Problems in Quantum Mechanics},
Pergamon, Oxford, (1971), ch. II.

\bibitem{symorse}
M. Znojil,
``Morse potential, symmetric Morse potential and bracketed
bound-state energies,''
Mod. Phys. Lett. {\bf A31} (2016)  1650088.

\bibitem{papone}
M. Znojil, ``Non-analytic exponential well $V(x)= -g^2\exp (-|x|)$ and an innovated, analytic
shooting method'', {\tt arXiv:1605.06780[quant-ph]}.


\bibitem{Ishkh1}
A. M. Ishkhanyan,
``Exact solution of the Schr\"{o}dinger equation for the inverse square root potential
$V_0/\sqrt{x}$,''  Eur. Phys.  Lett. {\bf 112} (2015) 10006.

\bibitem{afterIshkhanyan}
M.\, Znojil,
``Symmetrized exponential oscillator". Mod. Phys. Lett. A, to appear
{\tt arXiv:1609.00166[quant-ph].} 

\bibitem{Hille}
E. Hille, {\sl Ordinary Differential Equations in the Complex Domain},
Wiley, New York, (1976).

\bibitem{MAPLE}
https://www.maplesoft.com/products/maple/

\bibitem{watson}
G.\,N.\, Watson, {\sl A Treatise on the Theory of Bessel Functions},
Cambridge University Press 1922.


\bibitem{crum}
M.\,M.\,Crum, ``Associated Sturm-Liouville systems," Quart. J. Math.
Oxford Ser. (2) {\bf 6} (1955) 121-127, {\tt arXiv:physics/9908019}.

\bibitem{darboux}
G.\,Darboux, ``Sur une proposition relative aux \'equations
lin\'eaires." {\it C. R. Acad. Paris} {\bf 94} (1882) 1456-1459.

\bibitem{genden}
L.\,E.\,Gendenshtein, ``Derivation of exact spectra of the
Schroedinger equation by means of supersymmetry,'' JETP Lett. {\bf
38} (1983) 356-359.


\bibitem{KM}
I.\, Kay and H.\,M.\, Moses,
``Reflectionless transmission through dielectrics and scattering potentials,"
J. Appl. Phys. {\bf 27} (1956) 1503-1508.

\bibitem{refless}
R.\,Sasaki,
``Exactly solvable potentials with finitely many discrete eigenvalues of
arbitrary choice,'' J. Math. Phys. {\bf 55} (2014) 062101 (14pp),
{\tt arXiv:1402.5474\hspace{0pt}[math-ph]}.


\bibitem{erdelyi}
A. Erdelyi et al, {\sl Higher Transcendental Functions}, McGrawhill, New-York (1953), vol. 2, Ch. VII,
Bessel Functions, \S7.9 {\sl Zeros of Bessel Functions}, page 60.

\bibitem{coul}
M.\,J.\, Coulomb,
``Sur les z\'eros de fonctions de Bessel consid\'er\'ees comme fonctions de l'ordre,''
Bull. Sci. Math. {\bf 60} (1936) 297-302.

\bibitem{conde}
S.\,Conde and S.\,L.\,Kalla,
``The $\nu$-Zeros of $J_{-\nu}(x)$,''
Math. Computation {\bf 33} (1979) 423-426.


\end{thebibliography}
\end{document}